\documentclass[12pt]{article}

\usepackage{amsmath,amsthm,amssymb}
\usepackage{amsfonts,euscript}

\DeclareMathOperator{\diag}{diag}

\newcommand{\EL}{\EuScript L}
\newcommand{\EJ}{\EuScript J}
\newcommand{\EH}{\EuScript H}
\newcommand{\EP}{\EuScript P}
\newcommand{\prt}{\partial}
\newcommand{\ve}{\varepsilon}
\newcommand{\uuu}[1]{\overset{(3)}{#1}{}}
\newcommand{\ddd}[1]{\underset{\thicksim}{#1}{}}
\newcommand{\puasson}[2]{\left\{ #1, ~#2\right\}}
\newcommand{\dxx}{\delta^3(x-\tilde{x})}
\newcommand{\mae}{\varkappa}
\newcommand{\e}{\varepsilon}
\newcommand{\de}{\delta}
\newcommand{\dd}{\partial}
\newcommand{\too}{\longrightarrow}
\newcommand{\ls}{\left(}
\newcommand{\rs}{\right)}
\newcommand{\la}{\lambda}
\newcommand{\La}{\Lambda}

\makeatletter
\renewcommand{\section}{\@startsection{section}{1}{0pt}%
          {3.5ex plus 1ex minus .2ex}{2.3ex plus .2ex}{\noindent\hfil\bf}}
\makeatother

\begin{document}
\title{
Different canonical formulations\\
of Einstein's theory of gravity\\
}
\author{V.A.~Franke\thanks{E-mail: franke@pobox.spbu.ru}\\
{\it St.-Petersburg State University, Russia}
}
\date{\vskip 15mm}
\maketitle

\begin{abstract}
We describe the four most famous versions of the classical
canonical formalism in the Einstein theory of gravity: the
Arnovitt-Deser-Misner formalism, the Faddeev-Popov formalism, the
tetrad formalism in the usual form, and the tetrad formalism in the
form best suited for constructing the loop theory of gravity,
which is now being developed. We present the canonical
transformations relating these formalisms.
The paper is written mainly for pedagogical purposes.
\end{abstract}

\newpage
\section{Introduction}
The most direct method for constructing a quantum theory is to
quantize the corresponding classical theory written in canonical
form. Different equivalent canonical formulations of the classical
theory may then lead to not completely equivalent versions of the
quantum theory. In complicated cases, it is therefore beneficial
to use different methods to represent the classical theory in
canonical form before quantization. In particular, this concerns
the theory of gravity, whose final quantum form has not yet been
found. It is not improbable that choosing an appropriate classical
canonical formulation, we can here approach a satisfactory
solution of the quantization problem. Precisely this approach
underlies the so-called loop theory of gravity (see \cite{lg} and the
references therein), which is currently being developed.

In this paper written mainly for pedagogical purposes, we describe
several well-known equivalent classical canonical formulations of
the Einstein theory of gravity and relations between these
formulations. We first consider the Arnovitt-Deser-Misner (ADM)
formalism \cite{adm}. We then use the canonical transformation to pass to
the Faddeev-Popov (FP) formalism \cite{fp}. We next use a change of
variables to introduce the frame (tetrad) formalism in the usual form.
Finally, we use the canonical transformation to reduce this
formalism to the form underlying the loop theory of gravity \cite{lg}.

We do not consider the problem of quantizing gravity here and
restrict ourself to only several remarks on this subject, but the
information presented here can be useful in studying this problem.

\section{The ADM formalism}
First, we consider the classical ADM formalism \cite{adm}. Let $x^\mu$
be coordinates in the Riemannian space-time $(\mu,\nu,\ldots = 0, 1, 2, 3)$.
The coordinate $x^0=t$ is called time (we set $c=1$, where $c$
is the speed of light). We assume that all hypersurfaces $x^0=const$
are spacelike. The space coordinates are denoted
by $x^i$ ($i,k,\ldots=1,2,3$).
We use the metric signature $(-,+,+,+)$.

We fix
a hypersurface $x^0=const$ and let $\Sigma$ denote it. In the coordinates
$x^i$ the three-dimensional metric induced on $\Sigma$ coincides with the
three-dimensional part of the four-dimensio\-nal metric. We let $\beta_{ik}$
denote this three-dimensional metric and introduce $\beta^{ik}$ by the
condition
\begin{equation}
\beta_{ik}\beta^{kl}=\delta_{i}^{l}.
\end{equation}
Then
\begin{align}
\beta_{ik}=&~g_{ik},\\
\beta^{ik}=&~g^{ik}-\frac{g^{0i}g^{0k}}{g^{00}},
\end{align}
where $g_{\mu\nu}$ is the four-dimensional metric and
$g^{\mu\nu}g_{\nu\lambda}=\delta^{\mu}_\lambda$.
We introduce the notation
\begin{equation}
g=\det g_{\mu\nu},\qquad \beta=\det \beta_{ik}.
\end{equation}

As usual,
\begin{equation}
R^\alpha{}_{\beta,\gamma\delta}=\prt_\gamma
\Gamma^\alpha_{\delta\beta}-\prt_\delta
\Gamma^\alpha_{\gamma\beta}+\Gamma^\alpha_{\gamma\rho}\Gamma^\rho_{\delta\beta}-\Gamma^\alpha_{\delta\rho}\Gamma^\rho_{\gamma\beta},
\end{equation}
\begin{equation}
R_{\beta\delta}=R^\alpha{}_{\beta,\alpha\delta},
\end{equation}
\begin{equation}
R=g^{\beta\delta}R_{\beta\delta},
\end{equation}
where $\Gamma^\alpha_{\beta\gamma}$ are the Christoffel symbols
constructed from the metric $g_{\mu\nu}$ by a known method. We determine
the quantities $\uuu{\Gamma}{}^i_{kl}$, $\uuu{R}{}^i{}_{k,lm}$,
$\uuu{R}_{lm}$, $\uuu{R}$, formed
from $\beta_{ik}$,
$\prt_l\beta_{ik}$, $\prt_m\prt_l\beta_{ik}$ precisely as
the quantities $\Gamma^\alpha_{\beta\gamma}$, $R^\alpha{}_{\beta,\gamma\delta}$,
$R_{\beta\delta}$, $R$
are constructed from $g_{\mu\nu}$, $\prt_\alpha g_{\mu\nu}$,
$\prt_\alpha\prt_\beta g_{\mu\nu}$. We
introduce the covariant derivative $\uuu{\nabla}_i$ acting on $\Sigma$
by the
connection $\uuu{\Gamma}{}^i_{kl}$ just
as the derivative $\nabla_\mu$ acts on the entire
space-time by the connection $\Gamma^\alpha_{\beta\gamma}$.
We determine the second
fundamental tensor $K_{ik}$ of the hypersurface $\Sigma$:
\begin{equation}
K_{ik}=K_{ki}=\Bigl.-\nabla_i n_k\Bigr|_\Sigma=\Bigl. n_0
\Gamma^0_{ik}\Bigr|_\Sigma=\left.-\frac{1}{\sqrt{-g^{00}}}\Gamma^0_{ik}
\right|_\Sigma,
\end{equation}
where the field
$n_\mu(x)$ of unit normals to the surfaces $x^0=const$ is determined by
the relations
\begin{equation}\label{h3}
n_\mu n^\mu=-1,\qquad
n_\mu =-\delta^0_\mu\frac{1}{\sqrt{-g^{00}}},\qquad
n^\mu=-\frac{g^{0\mu}}{\sqrt{-g^{00}}}.
\end{equation}

We have the identity
\begin{equation}\label{kr4}
\sqrt{-g}\,R=\sqrt{-g}\left(\overset{(3)}{R}+K^i_l K^l_i-
(K^i_i)^2\right)+2\prt_\gamma\left(\sqrt{-g}(n^\gamma\nabla_\delta
n^\delta-n^\delta \nabla_\delta n^\gamma)\right),
\end{equation}
where $K_l^i=\beta^{ik}K_{kl}$. The
simplest derivation of this identity is based on the well-known
Gauss formula relating the curvature tensor of the hypersurface to
the curvature tensor of the ambient Riemannian space.

We consider
only the gravitational field not interacting with other fields
because all specific features of the problem can be clearly seen
in this case. We start from the action of the gravitational field
\begin{equation}\label{kkr4}
S=\int d^4 x\, \EL
\end{equation}
where
\begin{equation}\label{v4}
\EL=\frac{1}{2\mae}\sqrt{-g}\left(g^{\alpha\beta}\left(
\Gamma^\rho_{\alpha\gamma}\Gamma^\gamma_{\rho\beta}
-\Gamma^\rho_{\alpha\beta}\Gamma^\gamma_{\rho\gamma}\right) -
2\Lambda\right).
\end{equation}
Here $\mae= 8 \pi \gamma$, $\gamma$ is the Newtonian gravitational
constant, and $\Lambda$ is the cosmological constant.
Otherwise,
\begin{equation}
\EL=\frac{1}{2\mae} \sqrt{-g} \left( R-2\Lambda\right) +
\frac{1}{2\mae}\prt_\gamma \left( \sqrt{-g}\left(
g^{\gamma\alpha}
\Gamma^\beta_{\alpha\beta}-g^{\alpha\beta}\Gamma^\gamma_{\alpha\beta}
\right)\right).
\end{equation}
whence we
use identity \eqref{kr4} to obtain
\begin{multline}\label{kr5}
\EL=\frac{1}{2\mae} \sqrt{-g}\left(\overset{(3)}{R}+K^i_l
K^l_i- (K^i_i)^2-2\Lambda\right) +\\+
\frac{1}{2\mae}\prt_\gamma \left( \sqrt{-g}\left(
g^{\gamma\alpha}
\Gamma^\beta_{\alpha\beta}-g^{\alpha\beta}\Gamma^\gamma_{\alpha\beta}
\right)+
2\sqrt{-g}(n^\gamma\nabla_\delta
n^\delta-n^\delta \nabla_\delta n^\gamma)\right) .
\end{multline}
In the case of a closed universe, we can here
omit the divergence, and in the case of an island position of
masses in an asymptotically three-dimensionally flat space-time,
it suffices to only take into account the essential part of the
divergence equal to
\begin{equation}
\frac{1}{2\mae} \left( \prt_k \prt_k \beta_{ll} -
\prt_i\prt_k \beta_{ik}\right).
\end{equation}
In the last case $\Lambda=0$.
We often omit the divergence and assume that the
universe is closed for simplicity; we also often omit the $\Lambda$ term.

We choose the quantities
\begin{equation}
\beta_{ik}\equiv g_{ik},\qquad
N \equiv \frac{1}{\sqrt{-g^{00}}},\qquad
N_i=g_{0i}.
\end{equation}
as independent ADM field variables. In what follows, the subscripts
$i,k,\ldots$ are raised and lowered by the three-dimensional tensors
$\beta^{ik}$ and $\beta_{ik}$.
The following relations hold:
\begin{eqnarray}
&&g_{ik}=\beta_{ik},\qquad
g^{ik}=\beta^{ik}-\frac{N^i N^k}{N^2},\qquad
g_{0k}=N_k, \qquad
g^{0k}=\frac{N^k}{N^2},\\
&&g_{00}=-N^2+N_k N^k,\qquad g^{00}=-\frac{1}{N^2},\qquad
\sqrt{-g}=N\sqrt{\beta},\\
&&n_\mu=-\delta^0_\mu N,\qquad
n^0=\frac{1}{N},\qquad n^i=-\frac{N^i}{N},\\
&&K_{ik}=\frac{1}{2N}\left( \overset{(3)}{\nabla}_i
N_k+\uuu{\nabla}_k N_i - \prt_0 \beta_{ik} \right).\label{kr6}
\end{eqnarray}

In these
variables, Lagrangian density \eqref{kr5} with the divergence omitted
becomes
\begin{equation}
\EL^{(ADM)}=N \left\{ 2 \EJ^{ij,kl} K_{ij} K_{kl}
+\frac{\sqrt{\beta}}{2\mae}\left(\uuu{R}-2\Lambda\right)\right\},
\end{equation}
where
\begin{equation}
\EJ^{ij,kl}=\frac{1}{4}\left(\frac{\sqrt{\beta}}{2\mae}\right)
\left(
\beta^{ik}\beta^{jl}+\beta^{il}\beta^{jk}-2\beta^{ij}\beta^{kl}\right).
\end{equation}
It is convenient to introduce the symbols
\begin{equation}
\delta^{ik}_{lm} \equiv \frac{1}{2} \left( \delta^i_l
\delta^k_m+\delta^i_m \delta^k_l \right)
\end{equation}
and to determine the quantity $\EJ_{ij,kl}$ using the condition
\begin{equation}
\EJ^{ij,kl} \EJ_{kl,mn}=\delta^{ij}_{nm}.
\end{equation}
In this case, we have
\begin{equation}
\EJ_{ij,kl}=\left(\frac{2\mae}{\sqrt{\beta}}\right) \left(
\beta_{ik}\beta_{jl}+\beta_{il}\beta_{jk}-\beta_{ij}\beta_{kl}\right),\qquad
\EJ_{ij,kl}=\EJ_{kl,ij}=\EJ_{ji,kl}=\EJ_{ij,lk}
\end{equation}
and similar relations hold for $\EJ^{ij,kl}$.
We again omit the inessential part of the divergence arising in
the expression for $\EL$.

We set
\begin{equation}
L=\int\limits_{x^0=const} d^3 x ~\EL^{(ADM)}
\end{equation}
and
determine the conjugate momenta,
\begin{align}
&P^{(N)}(x)\equiv \frac{\delta L}{\delta (\prt_0 N(x))},\\
&P^{(N_i)}(x)\equiv \frac{\delta L}{\delta (\prt_0 N_i(x))},\\
&P^{ik}(x)\equiv \frac{\delta L}{\delta (\prt_0
\beta_{ik}(x))}=\frac{\prt \EL^{(ADM)}}{\prt (\prt_0
\beta_{ik}(x))},
\end{align}
where $x\equiv(x^1,x^2,x^3)$ and $\Bigl.\delta/\delta(\,)$ is the
three-dimensional variational derivative. We immediately obtain
the primary constraints
\begin{eqnarray}
P^{(N)}(x)=0,&& P^{(N_i)}(x)=0.
\end{eqnarray}
We solve
these constraints explicitly, i.~e., we set $P^{(N)}$ and $P^{(N_i)}$
equal to
zero everywhere as they are encountered.
Next,
\begin{equation}
P^{ik}(x)=\frac{\prt \EL^{(ADM)}}{\prt K_{lm}}\frac{\prt
K_{lm}}{\prt (\prt_0 \beta_{ik})}=-2 \EJ^{ik,lm}K_{lm},
\end{equation}
hence
\begin{equation}
K_{ik}=-\frac{1}{2}\EJ_{ik,lm}P^{lm},
\end{equation}
and according to \eqref{kr6}, we have
\begin{equation}
\prt_0 \beta_{ik}=\uuu{\nabla}_i N_k +\uuu{\nabla}_k N_i-2 N
K_{ik}.
\end{equation}

The density of the generalized Hamiltonian is equal to
\begin{equation}
\EH^{(gen)}=P^{ik}\prt_0 \beta_{ik} - \EL^{(ADM)}.
\end{equation}
We again omit the inessential
addition to the divergence and obtain
\begin{equation}
\EH^{(gen)}=N \EH_0+N^i\EH_i,
\end{equation}
where
\begin{equation}\label{sv1}
\EH_0=\frac{1}{2}\EJ_{ik,lm}P^{ik}P^{lm}+\left(\frac{\sqrt{\beta}}{2\mae}\right)
\left(-\uuu{R}+2\Lambda\right),
\end{equation}
\begin{equation}\label{sv2}
\EH_i=-2\beta_{is} \sqrt{\beta} \uuu{\nabla}_l \left(
\frac{P^{ls}}{\sqrt{\beta}}\right),
\end{equation}
and we take into account that $P^{ls}\bigr/\sqrt{\beta}$ is a tensor.

The density of the first-order Lagrangian is equal to
\begin{equation}
\EL^{(ADM)}_{(1)}=P^{ik} \prt_0 \beta_{ik} - \EH^{(gen)}=P^{ik}
\prt_0 \beta_{ik}-N \EH_0-N^i \EH_i.
\end{equation}
We add the essential part of the divergence and obtain the
relation for the island position of masses in asymptotically
three-dimensionally flat space:
\begin{equation}
\EL^{(ADM)}_{(1)}=P^{ik} \prt_0 \beta_{ik} - \EH^{(gen)}=P^{ik}
\prt_0 \beta_{ik}-N \EH_0-N^i \EH_i-\frac{1}{2\mae}\left(
\prt_i \prt_k \beta_{ik}- \prt_k \prt_k \beta_{ii} \right).
\end{equation}
We vary
$\EL^{(ADM)}_{(1)}$ in $N$ and $N^i$ and obtain the secondary
constraints
\begin{equation}
\EH_0(x)=0,\qquad \EH_i (x)=0.
\end{equation}
In the case of the island position of masses,
the total energy reduces to the surface integral
\begin{equation}
H^{\text{total}}=\frac{1}{2\mae} \int d^3 x \,\left( \prt_i
\prt_k \beta_{ik}- \prt_k \prt_k \beta_{ii} \right),
\end{equation}
and we now obtain
\begin{equation}
\EH^{(gen)}= N \EH_0+N^i\EH_i+ \frac{1}{2\mae}\left( \prt_i
\prt_k \beta_{ik}- \prt_k \prt_k \beta_{ii} \right).
\end{equation}
In the case of a closed
universe, the total energy is zero.

We introduce the Poisson
brackets. If $F_1$ and $F_2$ are two three-dimensional functionals of
$\beta_{ik}$ and $P^{ik}$, then
\begin{equation}
\left\{ F_1, F_2 \right\}=\int d^3 x \left( \frac{\delta
F_1}{\delta \beta_{ik}(x)} \frac{\delta F_2}{\delta
P^{ik}(x)}-\frac{\delta F_2}{\delta \beta_{ik}(x)} \frac{\delta
F_1}{\delta P^{ik}(x)} \right),
\end{equation}
where $\delta/\delta()$
is the three-dimensional variational derivative. Obviously,
\begin{equation}
\left\{ F_1, F_2 \right\}=-\left\{ F_2, F_1 \right\},
\end{equation}
\begin{equation}
\left\{ F_1, \left\{F_2 ,F_3\right\}\right\}+\left\{ F_2,
\left\{F_3 ,F_1\right\}\right\}+\left\{ F_3, \left\{F_1
,F_2\right\}\right\}=0,
\end{equation}
\begin{equation}
\left\{ F_1, F_2 F_3\right\}=\left\{ F_1, F_2 \right\}
F_3+F_2\left\{ F_1, F_3 \right\}.
\end{equation}
We next use the notation
\begin{eqnarray}
f\equiv f(x), && \ddd{f}\equiv f(\tilde x).
\end{eqnarray}
In this notation, we obtain
\begin{equation}
\puasson{\beta_{ik}}{\ddd{P}^{lm}}=\delta^{lm}_{ik}\,\delta^3(x-\tilde
x),
\end{equation}
\begin{eqnarray}
\puasson{\beta_{ik}}{\ddd{\beta}{}_{lm}}=0,&&\puasson{P^{ik}}{\ddd{P}{}^{lm}}=0.
\end{eqnarray}

The following relations hold:
\begin{equation}\label{kr10a}
\puasson{\EH_i}{\ddd{\EH}_k}=\EH_k \prt_i \delta^3(x-\tilde{x})-
\ddd{\EH}_i \ddd{\prt}_k \delta^3(x-\tilde{x}),
\end{equation}
\begin{equation}\label{kr10b}
\puasson{\EH_i}{\ddd{\EH}_0}=\EH_0 \prt_i \dxx,
\end{equation}
\begin{equation}\label{kr10c}
\puasson{\EH_0}{\ddd{\EH}_0}=\beta^{ik} \EH_k \prt_i \dxx -
\ddd{\beta}^{ik} \ddd{\EH}_k \ddd{\prt}_i \dxx,
\end{equation}
where $\ddd{\prt}_i=\frac{\prt}{\prt\tilde{x}^i}$. Clearly, all
the constraints in the classical theory are of the first kind. No
new constraints arise.

The constraints $\EH_i$ are generators of
three-dimensional transformations of coordinates on the surface $\Sigma$.
Indeed, after the change of coordinates
\begin{equation}
x^i\rightarrow x'\,^i+\xi^i(x),
\end{equation}
where $\xi^i(x)$ are infinitely small, we have
\begin{equation}
\delta \beta_{ik} \equiv \beta'_{ik}(x)-\beta_{ik}(x)
=-\uuu{\nabla}_i \xi_k-\uuu{\nabla}_k \xi_i,
\end{equation}
\begin{equation}
\delta P^{ik} \equiv P'\,^{ik}(x) - P^{ik}(x) = (\prt_l \xi^i)
P^{lk} + P^{il} \prt_l \xi^k - \prt_l (P^{ik} \xi^l).
\end{equation}
It can be verified directly that
\begin{equation}
\puasson{\int d^3x~\EH_i \xi^i}{\ddd{\beta}_{kl}}=\delta
\ddd{\beta}_{kl},\qquad
\puasson{\int d^3x~\EH_i \xi^i}{\ddd{P}^{kl}}=\delta \ddd{P}^{kl}.
\end{equation}
Correspondingly, the constraint $\EH_0$ generates displacements of
points of the surface $\Sigma$ along the normal to $\Sigma$.
In this case, the
variations in $\beta_{ik}$ and $P^{ik}$ correspond to the solutions of the
Einstein equations.

We make several remarks about quantizing the
described theory.
Under quantization, the variables $\beta_{ik}$ and $P^{ik}$
are replaced with operators satisfying the conditions
\begin{eqnarray}
&&[\beta_{ik},\ddd{P}^{lm}]=i\de_{ik}^{lm}\de^3(x-\tilde x),\\
&&[\beta_{ik},\ddd{\beta}_{lm}]=[P^{ik},\ddd{P}^{lm}]=0.
\end{eqnarray}
Because constraints \eqref{sv1} and \eqref{sv2} are too complicated to be
solved explicitly, these constraints are usually imposed on the
state vector. The theory thus obtained is consistent only under
the condition that the commutators of the constraints are equal to
linear combinations of these constraints with coefficients placed
to the left of them. After quantization, the constraints satisfy
commutation relations of form \eqref{kr10a}-\eqref{kr10c}
with the bracket $\{\,\,\,\}$
replaced with $-i[\,\,\,]$. But the order of the factors $\beta_{ik}$ and
$P^{ik}$ chosen in the expressions for the constraints is now
important. It may happen that the result of commuting the
constraints contains these factors not in the order originally
accepted in the constraints and the coefficients of the
constraints may arise not only to the left of them. It is easy to
see that this does not occur in quantum analogues of relations
\eqref{kr10a} and \eqref{kr10b},
and these relations preserve the form after
quantization (up to the change $\{\,\,\,\}\to-i[\,\,\,]$). In
particular, the latter is due to the abovementioned geometric
sense of the constraints $\EH_i$ as generators of transformations of
three-dimensional coordinates. This sense is completely preserved
under quantization.

The situation with the quantum analogue of
relation \eqref{kr10c} is quite different. If the operators in the
constraints $\EH_0$ and $\EH_i$ are located such that these constraints are
Hermitian, then the quantity $-i[\EH_0,\ddd{\EH}_0]$ obtained from the
quantity $\{\EH_0,\ddd{\EH}_0\}$ is also Hermitian. This means that the
non-Hermitian expressions $\beta^{ik}\EH_k$ cannot appear on the right in an
analogue of relation \eqref{kr10c} (we take into account that $\beta^{ik}$
and $\EH_k$ do
not commute). The most that can be obtained for the commutator
$-i[\EH_0,\ddd{\EH_0}]$ by choosing the order of the factors
in $\EH_0$ and $\EH_k$
without violating the Hermitian property is an expression of the
form
\begin{eqnarray}\label{zv}
&\frac{1}{2}\ls \beta^{ik}\EH_k+\EH_k\beta^{ki}\rs\dd_i
\de^3(x-\tilde x)-
\frac{1}{2}\ls \ddd{\beta}^{ik}\ddd{\EH}_k+
\ddd{\EH}_k\ddd{\beta}^{ki}\rs\ddd{\dd}_i\de^3(x-\tilde x)=\nonumber\\
&\hskip -8em=\beta^{ik}\EH_k\dd_i\de^3(x-\tilde x)-
\ddd{\beta}^{ik}\ddd{\EH}_k\ddd{\dd}_i\de^3(x-\tilde x)+\nonumber\\
&+\de^3(0)(\dots)\dd_i\de^3(x-\tilde x)-
\de^3(0)\ddd{(\dots)}\ddd{\dd}_i\de^3(x-\tilde x),
\end{eqnarray}
where $(\dots)$ and $\ddd{(\dots)}$ are some nonzero
operator-valued functions. The symbols $\de^3(0)$ arise from commuting
the operators $\beta^{ik}$ and $\EH_k$ or $\ddd{\beta}^{ik}$
and $\ddd{\EH}_k$ taken at the same
point.

Clearly, an expression of form \eqref{zv} containing the product
$\de^3(0)\dd_i\de^3(x-\tilde x)$ does not make sense.
A meaningful expression can
be obtained from it only by regularization. This raises the
question of the possibility of choosing a regularization such that
the extra terms in expression \eqref{zv} become zero and the general
covariance of the theory is reestablished after the regularization
is removed. But a unique answer to this question has not yet been
obtained. In several published works, the problem of
regularization and its removal was studied insufficiently
rigorously. An explanation for this is that the regularization
methods were studied in detail only in the framework of the
perturbation theory. But the problem is posed beyond this
framework here.

Although there is still a certain ambiguity in
this problem, the theory of gravity was quantized by the
path-integral method by analogy with quantizing non-Abelian gauge
theories (see \cite{fp} and the references therein
and also \cite{konpop}). If a
satisfactory perturbation theory were thus obtained, then its
consistency could be verified directly in the framework of the
Feynman diagram formalism, and this would suffice. But it turned
out that the constructed perturbation theory is unrenormalizable.
Under these circumstances, different approaches for constructing
the quantum theory of gravity are now being developed; the most
well-known approaches are superstring theory (see, e.g., \cite{grin}) and
the so-called loop theory of gravity \cite{lg}.

We also note that the
above difficulties in closing the constraint algebra after
quantization are also typical of other versions of the canonical
formalism in the theory of gravity, which are described below.

\section{The FP formalism}
We now consider the classical canonical FP
formalism \cite{fp}. We first introduce the quantities
\begin{equation}
h^{\mu\nu}=\sqrt{-g}\, g^{\mu\nu},
\end{equation}
in terms of which the subsidiary harmonic coordinate
condition can be simply written as $\prt_\mu h^{\mu\nu}=0$.
For the original
variables, we take the functions
\begin{equation}
q^{ik}\equiv h^{0i} h^{k0} - h^{00} h^{ik},
\end{equation}
and we write $q\equiv \det q^{ik}$ in what follows.
Moreover, we preserve the
functions $N$ and $N^i$ contained in the ADM formalism.

The ADM and FP
formalisms are related by the canonical transformation
\begin{equation}\label{kkr11}
 \left\{
\begin{aligned}
 &q^{ik}=\beta \beta^{ik} = \frac{1}{2} \ve^{ilm}
\ve^{knp} \beta_{ln} \beta_{mp},\\ &
\pi_{ik}=-\frac{1}{(2\mae)2\sqrt{\beta}}\EJ_{ik,lm}P^{lm}=\beta^{-1}\left(\frac{1}{2}\beta_{ik}\beta_{lm}
- \beta_{il} \beta_{km} \right) P^{lm},
\end{aligned}
\right.
\end{equation}
\begin{equation}\label{kr11}
 \left\{
\begin{aligned}
 &\beta_{ik}= \frac{1}{2\sqrt{q}} \ve_{ilm}
\ve_{knp} q^{ln} q^{mp},\\ &
P^{lm}=-\frac{1}{\sqrt{q}}\left(q^{ik}q^{lm} - q^{li} q^{mk}
\right) \pi_{ik},
\end{aligned}
\right.
\end{equation}
where $\pi_{ik}$ are the momenta conjugate to the generalized
coordinates $q^{ik}$, $\ve_{ikl}$ is a completely antisymmetric
symbol, and $\ve_{123}=1$. In this case
\begin{equation}
\pi_{ik} \prt_0 q^{ik}= P^{ik} \prt_0 \beta_{ik},
\end{equation}
\begin{equation}
\puasson{q^{ik}}{\ddd{\pi}_{lm}}=\delta^{ik}_{lm}\dxx,
\end{equation}
\begin{eqnarray}
\puasson{q^{ik}}{\ddd{q}^{lm}}=0, &&
\puasson{\pi_{ik}}{\ddd{\pi}_{lm}}=0.
\end{eqnarray}

In the FP formalism, the density of the first-order
Lagrangian has the form
\begin{equation}\label{kr12}
\EL_1^{\text{(FP)}}=\pi_{ik} \prt_0 q^{ik} - N \EH_0 - N^i \EH_i +
\frac{1}{2\mae} \prt_i \prt_k q^{ik},
\end{equation}
where we write the part of the divergence that is
essential in the case of the island position of masses in an
asymptotically three-dimensionally flat space-time. Now
\begin{equation}\label{kkr12}
\EH_0=\frac{2\mae}{q^{1/4}}\left( q^{lp} q^{mq} - q^{lm} q^{pq}
\right) \pi_{lm} \pi_{pq} -
\frac{q^{1/4}}{2\mae}\left(\uuu{R}-2\Lambda\right),
\end{equation}
\begin{equation}\label{v12}
\EH_i = \frac{2}{q^{1/4}} q^{kl}\left( \uuu{\nabla}_k \left(
q^{1/4} \pi_{il} \right) -\uuu{\nabla}_i \left( q^{1/4} \pi_{kl}
\right) \right),
\end{equation}
where we must
express $\beta_{ik}$ in $\uuu{R}$ in terms of $q^{lm}$ according to
\eqref{kr11}.

The quantities $\EH_0$ and $\EH_i$ continue to satisfy relations
\eqref{kr10a}-\eqref{kr10c}, and
the geometric meaning of these quantities is preserved.

\section{The usual frame formalism}
We now consider the frame formalism. At each
point of space-time, we introduce four mutually pseudo-orthogonal
normalized vectors $e_A^\mu (x)$, where the subscript $A$ numbers the
vectors $(A=0,1,2,3)$, the superscript $\mu$ numbers their
components in the coordinate basis $(\mu=0,1,2,3)$, and
$x\equiv \{ x^0,x^1,x^2,x^3 \}$.
We assume that
\begin{equation}\label{kr13}
e^\mu_A(x) g_{\mu\nu} e^\nu_B (x) = \eta_{AB},
\end{equation}
where $\eta_{AB}=\diag (-1,1,1,1)$. We also introduce
the variables $e^A_\mu (x)$
by the relation
\begin{equation}\label{kkr13}
e^A_\mu e^\nu_A = \delta^\nu_\mu.
\end{equation}
It follows from
relations \eqref{kr13} and \eqref{kkr13} that
\begin{equation}
g_{\mu\nu}(x) = e^A_\mu \eta_{AB} e^B_\nu.
\end{equation}
We substitute this expression in
expressions \eqref{kkr4} and \eqref{v4} above for
the action of the gravitational field and regard $e^A_\mu(x)$ as
functions describing this field in what follows. We thus obtain a
theory invariant under two groups of transformations: general
coordinate transformations and local Lorentz transformations of
the variables $e^A_\mu(x)$. The last transformations have the form
\begin{equation}\label{h13}
e'\,^A_\mu(x) = \omega^A{}_B(x)e^B_\mu(x),
\end{equation}
under the condition
\begin{equation}\label{v13}
\eta_{AB}\omega^A{}_D(x) \omega^B{}_E(x)=\eta_{DE}.
\end{equation}
Relation \eqref{v13} ensures the invariance of the metric
$g_{\mu\nu}(x)$ under such transformations. The variables $e^A_\mu$
and $e^\mu_A$ are
called frame parameters.

Each vector referred to the coordinate
basis is assigned a vector referred to the frame basis according
to the rule
\begin{eqnarray}
a^A=e^A_\mu a^\mu, && a_A=e^\mu_A a_\mu.
\end{eqnarray}
The tensors $T^{\mu,\ldots,A,\ldots}_{\nu,\ldots,B,\ldots}$,
which vary in the indices $\mu,\dots,\nu,\ldots$
as usual under the change of coordinates and which are
transformed by the Lorentz matrices
in the indices $A,\dots,B,\ldots$ under the change of parameters $e^A_\mu$
according to \eqref{h13}, can be
introduced similarly. The tensor indices $A,\dots,B,\ldots$ are
raised and lowered using the symbols $\eta^{AB}$ and $\eta_{AB}$.

The covariant
derivatives are introduced as usual,
\begin{eqnarray}
\nabla_\mu a^\nu= \prt_\mu a^\nu +\Gamma^\nu_{\mu\alpha}a^\alpha,
&& \nabla_\mu a_\nu= \prt_\mu a_\nu - a_\alpha
\Gamma^\alpha_{\mu\nu},
\end{eqnarray}
\begin{eqnarray}
\nabla_\mu a^A= \prt_\mu a^A +A_\mu{}^A{}_B a^B, && \nabla_\mu
a_A= \prt_\mu a_A - a_B A_\mu{}^B{}_A
\end{eqnarray}
(and similarly in the case of tensors). Under the
assumption that
\begin{eqnarray}
\nabla_\mu e^A_\nu=0, &  \nabla_\mu e_A^\nu=0, & \nabla_\mu
\eta^{AB}=0,
\end{eqnarray}
we establish the relation between $\Gamma^\alpha_{\mu\nu}$
and $A_\mu{}^A{}_B$:
\begin{equation}
\Gamma^\mu_{\alpha\nu}=e^\mu_A A_\alpha{}^A{}_B e^B_\nu + e^\mu_A
\prt_\alpha e^A_\nu,
\end{equation}
\begin{equation}\label{kr14}
A_\alpha{}^A{}_B =e^A_\mu \Gamma^\mu_{\alpha\nu} e^\nu_B + e^A_\mu
\prt_\alpha e^\mu_B,
\end{equation}
where $A_\alpha{}^A{}_B$ is called a frame connection and
$\Gamma^\mu_{\alpha\nu}$ is called a
coordinate connection.

The frame connection is similar to the
gauge field with a Lorentz structure group. Therefore,
\begin{equation}
A_\alpha{}^A{}_B=A_\alpha{}^{AD}\eta_{DB},
\end{equation}
where $A_\alpha{}^{AD} = - A_\alpha{}^{DA}$. If $A_\alpha$
is understood as the
matrix $A_\alpha{}^A{}_B$, then we can construct the analogue of the field
strength
\begin{equation}
F_{\mu\nu}=\prt_\mu A_\nu - \prt_\nu A_\mu + A_\mu A_\nu - A_\nu
A_\mu,
\end{equation}
and, moreover,
\begin{equation}
R^\alpha{}_{\beta,\mu\nu}=e^{\alpha}_A
F_{\mu\nu,}{}^A{}_B e^B_\beta.
\end{equation}

It is necessary to use the frame
formalism to describe spinors in the Riemannian space-time because
the spinor representations of the Lorentz group cannot be extended
to the representations of the total linear group. Therefore, the
spinors cannot be referred to the local coordinate basis; they can
only be referred to the pseudo-orthogonal frame basis. But we use
the frame formalism for a different purpose in what follows. As
before, we assume that the gravitational field does not interact
with other fields.

To remove the gauge arbitrariness completely,
we must additionally impose four coordinate and six frame
subsidiary conditions. We use this possibility and remove only
part of the frame arbitrariness using the three conditions
\begin{equation}\label{kr15}
e^\mu_{(0)}(x)=n^\mu,
\end{equation}
where $n^{\mu}(x)$ is the normalized normal to the
surface $x^0=const$ at the point $x$ (see relations \eqref{h3}).
Hereafter,
we enclose the frame indices in parentheses if they are written as
numbers and write the coordinate indices without parentheses
in this case. Relation \eqref{kr15} contains only three conditions because
the vectors $n^\mu$ and $e^\mu_{(0)}$ are normalized. After conditions
\eqref{kr15} are introduced,
the theory remains invariant under the semidirect
product of the group of general coordinate transformations and the
group of three-dimensional orthogonal transformations (i.~e., the
Lorentz transformations under which
the vector $e^\mu_{(0)}(x)=n^\mu (x)$
remains unchanged).

On each surface $x^0=const$, we introduce the
ADM variables $\beta_{ik}\equiv g_{ik}$, $N$ and $N^i$.
Next, $i,k,\dots$ are
three-dimensional coordinate indices $(i,k,\dots{} =1,2,3)$, and
$a,b,\dots$ are three-dimensional frame indices $(a,b,\dots{}=1,2,3)$.
By conditions \eqref{kr15}, the frame parameters $e^\mu_A$ and $e^A_\mu$ can
be expressed in terms of $e^i_a$, $e^a_i$, $N$ and $N^i$ as
\begin{equation}
e^0_{(0)}=\frac{1}{N},\qquad
e^0_a=0,\qquad
e^i_{(0)}=-\frac{N^i}{N},\qquad
e^{(0)}_0 = N,\qquad
e^{(0)}_i=0,\qquad
e^a_0=e^a_i N^i.
\end{equation}
In this case, not only
\begin{equation}
e_\mu^A e^\nu_A = \delta^\nu_\mu,
\end{equation}
but also
\begin{equation}\label{kr16}
e_i^a e^k_a = \delta^i_k,\qquad
g_{ik} = \beta_{ik} = e^a_i e^a_k,\qquad
\beta^{ik}=e_a^i e_a^k.
\end{equation}
We set
\begin{equation}
e \equiv \det e^a_i,
\end{equation}
and then
\begin{equation}
\beta= e^2.
\end{equation}

Having in mind a possible
application to the loop theory of gravity, for the main variables,
we take the functions
\begin{equation}\label{kr17}
Q^i_a\equiv e e^i_a = \sqrt{\beta} e^i_a, \qquad N,\qquad N_i.
\end{equation}
We set
\begin{equation}
Q \equiv \det Q^i_a,
\end{equation}
and
\begin{equation}\label{vv17}
Q= \beta.
\end{equation}
We define the quantities $Q^a_i$ by
the relations
\begin{equation}\label{kkr17}
Q^a_i Q_a^k = \delta_i^k.
\end{equation}
The vectors $Q^i_a$ (as well as
$e^i_a$) are tangent to the surface $x^0 = const$. It follows from
relations \eqref{kr16} that the indices $a,b, \ldots$ can be raised and
lowered using the symbols $\delta^{ab}$ and $\delta^{ab}$.
Therefore, there is no
difference between the superscripts and subscripts $a$, $b$, and
they can be written for convenience.

We develop the canonical
formalism on the hypersurface $x^0=const$ in terms of the
three-dimensional variables $Q^i_a$, $N$, $N^i$, preserving the
notation $\Sigma$ for this hypersurface. The simplest way to the goal is
to start from the FP formalism (see Sec.~3). According to formulas
\eqref{kkr11}, \eqref{kr16} and \eqref{kr17},
we have
\begin{equation}\label{v17}
q^{ik}=\beta \beta^{ik} = \beta e^i_a e^k_a = Q^i_a Q^k_a.
\end{equation}
We substitute this expression in FP Lagrangian \eqref{kr12} and first
assume that $\pi_{ik}$ and $Q^i_a$ are independent.
We see that
\begin{equation}
\pi_{ik} \prt_0 q^{ik} = 2 \pi_{ik} Q^k_a \prt_0 Q^i_a,
\end{equation}
and the variables $Q^i_a$ are thus assigned
the conjugate momenta
\begin{equation}
\EP^a_i=2 \pi_{ik} Q^k_a.
\end{equation}
We hence have
\begin{equation}\label{kkkr18}
\pi_{ik}=\frac{1}{2} Q^a_k \EP^a_i.
\end{equation}
But $\pi_{ik}=\pi_{ki}$, and the new constraints
\begin{equation}
Q^a_k \EP^a_i - Q^a_i \EP^a_k =0
\end{equation}
therefore appear. In view of relations \eqref{kkr17}, this
is equivalent to the three constraints
\begin{equation}\label{kr18}
\Phi^a \equiv \ve^{abc} Q^{ib} \EP_i^c =0,
\end{equation}
where $\ve^{abc}$ is completely antisymmetric and $\ve^{123}=1$.

The action of
the frame formalism can now be written in the canonical form
\begin{equation}
S^{(\text{frame})}_{(1)}=\int d^4 x\, \EL^{(\text{frame})}_{(1)},
\end{equation}
\begin{equation}\label{v18}
\EL^{(\text{frame})}_{(1)}=\EP^a_i \prt_0 Q^i_a - N \EH_0 - N^i
\EH_i - \lambda^a \Phi^a,
\end{equation}
where we take the new constraints into account using the Lagrange
multipliers $\lambda^a$. We assume that in FP formulas
\eqref{kkr12} and \eqref{v12} for $\EH_0$ and $\EH_i$, the
variables $q^{ik}$ and $\pi_{ik}$ are expressed in terms of
$\EP_i^a$ and $Q^i_a$ a according to \eqref{v17} and
\eqref{kkkr18}. Hereafter, for simplicity, we do not write the
divergence in $\EL^{(\text{frame})}_{(1)}$ and consider the case
of a closed universe. We have
\begin{equation}\label{h18}
\EH_0 = \frac{1}{4} \left( \frac{2\mae}{\sqrt{Q}}\right) \left(
Q^k_b Q^l_b \EP^c_k \EP_l^c - (Q^k_b \EP^b_k)^2 \right) - \left(
\frac{\sqrt{Q}}{2\mae} \right) \left( \uuu{R} - 2\Lambda\right),
\end{equation}
\begin{equation}\label{vv18}
\EH_i = Q^k_a \left( \uuu{\nabla}_k \EP^a_i - \uuu{\nabla}_i
\EP^a_k \right),
\end{equation}
where $\uuu{\nabla}_k$ as before is the covariant derivative on
the hypersurface $x^0 = const$ containing the connection coefficients
$\uuu{\Gamma}^i_{kl}$ and $\uuu{A}_i{}^a{}_b$
expressed in terms of $Q_a^k$ and $Q^a_k$.

The coefficients $\uuu{A}_i{}^a{}_b$ are determined
by analogy with formula
\eqref{kr14} by the relations
\begin{equation}\label{kr19}
\uuu{A}_k{}^a{}_b = e^a_i \uuu{\Gamma}^i_{kl} e^l_b - e^a_i \prt_k
e^i_b.
\end{equation}
We can verify that by condition \eqref{kr15}
\begin{equation}\label{v19}
\uuu{A}_k{}^a{}_b = A_k{}^a{}_b,
\end{equation}
where $A_k{}^a{}_b$ is the three-dimensional part of the
four-dimensional connection $A_\mu{}^A{}_B$. The quantities $A_k{}^a{}_b$
form an
$SO(3)$ connection. Therefore,
\begin{equation}
A_k{}^a{}_b=A_k{}^{ab}=-A_k{}^{ba}=\ve^{abc} A_k^c,
\end{equation}
where
\begin{equation}\label{h19}
A^c_k=\frac{1}{2} \ve^{cab} A_k{}^{ab}.
\end{equation}
It follows from
formulas \eqref{kr17} and \eqref{vv17} that
\begin{eqnarray}
e^i_a= Q^{-1/2} Q^i_a, && e^a_i = Q^{1/2} Q^a_i.
\end{eqnarray}
Substituting this in relation \eqref{kr19} and taking expressions
\eqref{v19} and \eqref{h19} into account, we obtain
\begin{multline}\label{kkr19}
A^c_i = \frac{1}{2} \ve^{cab} \left( (\prt_k Q^a_i -\prt_i Q^a_k)
Q^k_b - Q^l_a (\prt_l Q^d_m) Q^m_b Q^d_i + Q^b_i Q^k_a Q^l_d
\prt_k Q^d_l \right) = \\ = \frac{1}{2} \ve^{cab} \left( Q^a_k
\prt_i Q^{kb} + Q^d_i Q^{ka} (Q^b_l \prt_k Q^{ld} + Q^d_l \prt_k
Q^{lb}) + Q^a_i Q^{kb} Q^d_l \prt_k Q^{ld} \right).
\end{multline}

Letting $A_k$ denote the matrix $A_k{}^{ab}$, we can determine
the three-dimensional field strength
\begin{equation}
\uuu{F}_{ik} = \prt_i A_k - \prt_k A_i + A_i A_k - A_k A_i.
\end{equation}
In this case, the relations
\begin{equation}
\uuu{F}_{ik}{}^{ab}=\ve^{abc} \uuu{F}_{ik}{}^c,\qquad
\uuu{R}^i_{k,lm}= e^i_a \uuu{F}_{lm}{}^{ab} e^b_k
\end{equation}
hold.

The canonical
variables now satisfy the relations
\begin{equation}
\puasson{Q^i_a}{\ddd{\EP}^b_k}=\delta^i_k \delta^a_b \dxx,\qquad
\puasson{Q^i_a}{\ddd{Q}^k_b} = 0,\qquad
\puasson{\EP^b_i}{\ddd{\EP}^a_k} = 0.
\end{equation}
The symmetry condition $\pi_{ik} = \pi_{ki}$
in the framework of the formalism
considered in this section is satisfied because of constraints
\eqref{kr18}, while this condition holds in Sec.~3 by definition.
The Poisson brackets relating $\pi_{ik}$ and $\ddd{\pi}_{lm}$
differ from those introduced in Sec.~3.
We now have
\begin{eqnarray}
\puasson{q^{ik}}{\ddd{q}^{lm}} = 0, &&
\puasson{q^{ik}}{\ddd{\pi}_{lm}} = \delta^{ik}_{lm} \dxx,
\end{eqnarray}
but
\begin{equation}
\puasson{\pi_{ik}}{\ddd{\pi}_{lm}} = \frac{1}{4} Q^b_k Q^b_m Q^d_i
Q^c_l \ve^{cda} \Phi^a \dxx.
\end{equation}

New terms therefore appear in the
right-hand sides of relations with
Poisson brackets \eqref{kr10a}-\eqref{kr10c}, but
all these terms are proportional to the constraints $\Phi^a$.
Moreover,
\begin{equation}
\puasson{\Phi^a}{\ddd{\Phi}^b}= \ve^{abc} \Phi^c \dxx,\qquad
\puasson{\Phi^a}{\ddd{\EH}_0} = 0,\qquad
\puasson{\Phi^a}{\ddd{\EH}_i} = 0.
\end{equation}
Therefore, the classical algebra of constraints is closed,
and all the constraints $\EH_0$, $\EH_i$ and $\Phi^a$
are constraints of the first kind.

Instead of the constraints $\EH_i$ it is convenient to introduce
the constraints
\begin{equation}
\EH_i ' \equiv \EH_i + A_i{}^c \Phi^c \equiv Q^{bk} (\prt_k
\EP^b_i - \prt_i \EP^b_k ) + (\prt_k Q^{kb} ) \EP^b_i = 0.
\end{equation}
We can verify that the quantities $\EH_i '$
generate transformations of three-dimensional coordinates without
changing the frames as geometric objects and the quantities $\Phi^a$
generate rotations of frames without changing the coordinates.

The algebra of constraints in terms of the quantities $\EH_0$, $\EH_i '$
and $\Phi^a$ has the form
\begin{eqnarray}
&&\puasson{\EH_i '}{\ddd{\EH}_k '}=\EH_k ' \prt_i \dxx -
\ddd{\EH}_i' \ddd{\prt}_k \dxx,\nonumber \\
&&\puasson{\EH_i '}{\ddd{\EH}_0}= \EH_0 \prt_i \dxx,\nonumber \\
&&\puasson{\EH_0}{\ddd{\EH}_0}= Q^i_a Q^k_a \EH_k ' \prt_i \dxx -
\ddd{Q}^i_a \ddd{Q}^k_a \ddd{\EH}_i ' \ddd{\prt}_k \dxx +
(\ldots)^a \Phi^a - \ddd{(\ldots)}^a \ddd{\Phi}^a,\nonumber \\
&&\puasson{\Phi^a}{\ddd{\Phi}^b} = \ve^{abc} \Phi^c \dxx,\nonumber \\
&&\puasson{\Phi^b}{\ddd{\EH}_i '} = - \Phi^b \prt_i \dxx,\nonumber \\
&&\puasson{\Phi^a}{\ddd{\EH}_0} = 0,
\end{eqnarray}
where $(\ldots)^a$ are certain
expressions composed of $Q^i_a$ and $\EP_i^a$.

\section{The formalism used in the loop theory of gravity}
We now turn to the canonical formalism
underlying the loop theory of gravity. We can readily verify that
three-dimensional frame connection \eqref{kkr19} admits the representation
\begin{equation}\label{kr21}
A^a_i=\frac{\de F}{\de Q_a^i(x)},
\end{equation}
where
\begin{equation}
F=\frac{1}{2}\int\limits_{x^0=const}\!\! d^3 x \,\e^{abc} Q_a^lQ_k^b\dd_l Q^{kc}.
\end{equation}
Here, $x\equiv (x^1,x^2,x^3)$, and $\de/\de Q^i_a$ a is the
three-dimensional variational derivative. Because
\begin{equation}
\puasson{\EP^a_i(x)}{f[Q^k_b]}=-\frac{\de f[Q^k_b]}{\de Q^i_a(x)},
\end{equation}
where $f[Q^k_b]$ is an arbitrary
functional of the functions $Q^k_a(x)$, we have
\begin{equation}\label{kr22}
\puasson{\EP_m^f}{\ddd{A}_i^c}=\puasson{\ddd{\EP}_i^c}{A_m^f}.
\end{equation}
This permits performing the canonical
transformation
\begin{eqnarray}\label{kkr22}
Q_a^i&\too& Q_a^i,\nonumber\\
\EP_i^a&\too& \EP'^a_i=\EP_i^a+bA_i^a,
\end{eqnarray}
where $b$ is a number called the Barbero-Immirzi parameter. This
parameter can be assigned any value. By \eqref{kr22}, the condition
\begin{equation}
\puasson{\EP'^a_i}{\EP'^b_k}=0.
\end{equation}
holds. Because $A^a_i$ depends only on $Q^a_i$, the
relation
\begin{equation}
\puasson{Q^i_a}{\ddd{\EP}'^b_k}=\de^i_k\de^b_a\,\de^3(x-\tilde x)
\end{equation}
is also
satisfied.

After \eqref{kkr22}, we can perform one more canonical
change of variables:
\begin{eqnarray}\label{v22}
Q_a^i&\too&B^a_i=\frac{1}{b}\EP'^a_i=A^a_i+\frac{1}{b}\EP^a_i,\nonumber\\
\EP'^a_i&\too& \Pi^i_a=-bQ^i_a.
\end{eqnarray}

Under the change of the
three-dimensional coordinates on the hypersurface $x^0=const$ and
of the frames satisfying condition \eqref{kr15} all the time, the quantity
$B^a_i$ transforms as the frame connection $A^a_i$. Therefore, the path
integral
\begin{equation}
{\rm\bf tr}\,W(C)={\rm\bf tr}P\exp\ls
-\oint\limits_C dx^i\,B_i\rs,
\end{equation}
where $B_i$ is the matrix with the entries $B_i^{ab}=\e^{abc}B_i^c$
and the integration is over a closed path on the
hypersurface $x^0=const$, is invariant under the $SO(3)$
frame transformations generated by the constraints
$\Phi^a$.
This fact underlies the loop theory of gravity in which the quantities
$B^a_i$ and $\Pi^i_a$ are used as canonical variables.

It follows from formulas \eqref{v22} that
\begin{equation}
Q_a^i=-\frac{1}{b}\Pi_a^i,\qquad
\EP_i^a=b\ls B_i^a-A_i^a\rs,
\end{equation}
where
\begin{equation}
A^a_i\equiv A^a_i\Bigr|_{Q_a^i=-b^{-1}\Pi^i_a}.
\end{equation}
We substitute this expression in action \eqref{v18}, take \eqref{h18} and
\eqref{vv18} into account, and write the result in the two forms
\begin{equation}
S_1=\int d^4x\,\, \EL_1,
\end{equation}
\begin{eqnarray}
\EL_1
=&\Pi_a^i\dd_0B_i^a-N'\EH'_0-N'^i\EH'_i-\la'^a\Phi^a=\nonumber\\
=&\Pi_a^i\dd_0B_i^a-N''\EH''_0-N''^i\EH''_i-\la''^a\Phi^a,
\end{eqnarray}
where
\begin{equation}
\Phi^a=-\ls \dd_i\Pi_a^i+\e^{abc}B_i^c\Pi^i_b\rs=
-\sqrt{-\Pi}\,\overset{(3B)}{\nabla}_{\!\!i}\ls\frac{\Pi_a^i}{\sqrt{-\Pi}}\rs\equiv -D_i\Pi_a^i,
\end{equation}
\begin{eqnarray}
&&\EH'_k=\EH_k+A_k^a\Phi^a=-\Pi_c^i\uuu{F}^c_{ik}(B)+B_k^b\Phi^b,\nonumber\\
&&\EH'_0=\Pi^i_a\Pi^k_b\e^{abc}\uuu{F}_{ik}^c(B)+
4b^{-3}\frac{1}{(2\mae)^2}\Pi\ls(1+\mae^2b^2)\uuu{R}-2\La\rs,
\end{eqnarray}
\begin{eqnarray}
&\EH''_k=&\!\!\!-\Pi_c^i\uuu{F}^c_{ik}(B)+B_k^b\Phi^b,\hfill\nonumber\\
&\EH''_0=&\!\!\!\Pi^i_a\Pi^k_b\e^{abc}\uuu{F}_{ik}^c(B)-\nonumber\\
&&\!\!\!\!\!-(1+\mae^2b^2)\!\ls\Pi^k_b\Pi^l_b(B_k^c-A_k^c)(B_l^c-A_l^c)-
\ls\Pi_b^k(B^b_k-A^b_k)\rs^2\rs-2b^{-1}\La\Pi.
\end{eqnarray}
Here,
\begin{eqnarray}
&&N'=\frac{N}{4}\ls\frac{2\mae}{\sqrt{Q}}\rs,\qquad N'^k=N^k,\nonumber\\
&&\la'^a=\la^a+\frac{N}{4}\ls\frac{2\mae}{\sqrt{Q}}\rs
Q^{kd}\EP_k^b\e^{dba}-N^kA_k^b+b\mae e^i_a\dd_iN,\nonumber\\
&&N''=-\frac{N}{(2\mae)b^2\sqrt{Q}},\qquad N''_k=N_k,\nonumber\\
&&\la''^a=\la^a-N^kA_k^b-N\ls(2\mae)b^2\sqrt{Q}\rs^{-1}
Q^k_d\EP^f_k\e^{dfa}-2\ls(2\mae)b\rs^{-1}e^i_a\dd_iN,\nonumber\\
&&\Pi=\det \Pi^i_a,\qquad A^a_k=A^a_k(Q^i_a)
\Bigr|_{Q^i_a=-b^{-1}\Pi^i_a},\qquad
\uuu{R}=\uuu{R}(Q^i_b)\Bigr|_{Q^i_b=-b^{-1}\Pi^i_b},\nonumber\\
&&F_{ik}^{(3)}(B)=\dd_iB_k-\dd_kB_i+B_iB_k-B_kB_i,
\end{eqnarray}
$B_i$, $\uuu{F}_{ik}$ are matrices with the entries
\begin{equation}
B_i^{ab}=\e^{abc}B_i^c,\qquad \uuu{F}^{ab}_{ik}=\e^{abc}\uuu{F}^c_{ik},
\end{equation}
and $\overset{(3B)}{\nabla}_{\!\!i}$ is the three-dimensional
covariant derivative with the connection $B^{ab}_i$.

We can see
that the expressions for $\EH'_0$ and $\EH''_0$ are drastically simplified
and become equal to each other if we continue the theory to the
complex domain and set
\begin{equation}
b=\mp i\mae.
\end{equation}
According to \eqref{v22}, we then
have
\begin{equation}\label{kr25}
B_i^c=A_i^c\pm i\mae\EP^c_i.
\end{equation}
But in the frame basis where $e^\mu_{(0)}=n^\mu$,
the relations
\begin{equation}
A_i^{(0)c}=N\Gamma_{ik}^0 e^k_c=-K_{ik}e^k_c,
\end{equation}
hold, and then
\begin{equation}
\pi_{ik}=\frac{1}{(2\mae)\sqrt{\beta}}K_{ik},\qquad
\EP^a_i=2\pi_{ik}Q^k_a=\frac{1}{\mae}K_{ik}e^k_a=-\frac{1}{\mae}A_i^{(0)c},
\end{equation}
which implies that relation \eqref{kr25} becomes
\begin{equation}
B_i^c=A_i^c\mp iA_i^{(0)c},
\end{equation}
i.~e.,
\begin{equation}
B^{ab}_i=A^{ab}_i\mp i\e^{abc}A_i^{(0)c}.
\end{equation}

We can introduce the fields
\begin{equation}
A_\mu^{AB(\pm)}=A_\mu^{AB}\mp\frac{i}{2}\eta^{AD}\eta^{BE}\e_{DEFH}A_\mu^{FH},
\end{equation}
whence
\begin{equation}
A^{ab(\pm)}_i=A^{ab}_i\mp i\e^{abc}A_i^{(0)c}.
\end{equation}
The fields $A_\mu^{AB(\pm)}$ satisfy the self-duality (anti-self-duality)
conditions
\begin{equation}
A^{AB(\pm)}_\mu=\mp\frac{i}{2}\eta^{AD}\eta^{BE}\e_{DEFH}A_\mu^{FH(\pm)},
\end{equation}
and
\begin{equation}
A_\mu^{AB}=A_\mu^{AB(+)}+A_\mu^{AB(-)}.
\end{equation}
Therefore, the expressions for $\EH'_0$ and
$\EH''_0$ are simplified if
\begin{equation}
B_\mu^{ab}=A_\mu^{ab(+)}
\end{equation}
or
\begin{equation}
B_\mu^{ab}=A_\mu^{ab(-)}.
\end{equation}
In this case, all the constraints depend polynomially on
the variables $B_i^a$ and $\Pi_a^i$. But to return to the real domain, the
complicated condition
\begin{equation}\label{kr26}
B_i^a+{B_i^a}^*=2A(Q)\Bigr|_{Q_a^i=-b^{-1}\Pi^i_a}
\end{equation}
must be satisfied, where $*$ denotes complex conjugation in the
classical case and Hermitian conjugation after quantization.
Because condition \eqref{kr26} is complicated, it is currently preferred
to construct the loop theory of gravity for a real value of the
parameter $b$ for the case in which the constraint $\EH'_0$ (or $\EH''_0$)
is very complicated.

The above classical canonical formulations of
the theory of gravity have found and continue to find application
in studying the problem of quantizing this theory.

\vskip 1em
{\bf Acknowledgments.} The author thanks the UNESCO Regional Bureau for
Science and Culture in Europe for the support of the V.A.~Fock
International School of Physics.

\end{document}